\documentclass[aps,pre,reprint,groupedaddress]{revtex4-2}
\usepackage{amsmath}
\usepackage{siunitx}
\usepackage{color}
\usepackage{hyperref}
\usepackage{graphicx}
\usepackage{appendix}
\usepackage{xeCJK}
\begin{document}

\title{Competing Symmetry Breakings and Spontaneous Parity Restoration in Desynchronously Pumped Kerr Resonators}

\author{ Jun Ji$^{1,2}$}
\author{ Zhibin Ren$^{1,2}$}
\author{ Benli Yu$^{1,2}$}
\author{ Zhiqiang Wang$^{1,2}$}
\email[]{Contact author: zhiqiangwang@ahu.edu.cn}
\affiliation{$^1$State Key Laboratory of Opto-Electronic Information Acquisition and Protection Technology $\&$ School of Optoelectronic \\ Science and Engineering, Anhui University, Hefei, 230601, Anhui, P. R. China\\
$^2$Information Materials and Intelligent Sensing Laboratory of Anhui Province, Anhui University, Hefei 230039, China}

\date{\today}

\begin{abstract}
Symmetry breaking is a cornerstone of nonlinear science, where stronger driving typically compounds parity loss across coupled degrees of freedom. Here, we uncover a counterintuitive symmetry-restoring transition driven by the competition between internal spontaneous polarization symmetry breaking (PSB) and external explicit temporal symmetry breaking (TSB) in desynchronously pumped Kerr resonators. Tuning pump-cavity desynchronization steers dissipative cavity solitons from pure PSB, through a robust hybrid symmetry-breaking (HSB) regime where both asymmetries coexist, into a pure TSB state accompanied by the abrupt, spontaneous restoration of polarization parity. Using collective-coordinate perturbation theory, we map the effective Ginzburg-Landau potential landscape and show that kinematic drift weakens local driving until a saddle-node bifurcation eliminates the temporal confining barrier. Corroborated by full Ikeda-map simulations across the phase boundary, our results demonstrate how driven-dissipative systems sacrifice internal order to satisfy external kinematic constraints, offering a general paradigm for competing order parameters in multidimensional nonlinear systems.
\end{abstract}

%\keywords{}

\maketitle

The spontaneous breaking of fundamental symmetries is a hallmark of self-organization in non-equilibrium thermodynamic systems \cite{cross1993pattern}. In nonlinear optics, this principle is elegantly realized by dissipative structures \cite{grelu2012dissipative}, particularly Kerr cavity solitons (CSs) \cite{leo2010temporal,herr2014temporal,lilienfein2019temporal}. Sustained by a double balance of nonlinearity against dispersion, and parametric gain against cavity loss, this self-organization drives the spontaneous localization of a continuous-wave (CW) background into ultrashort pulses. While their corresponding optical frequency combs have revolutionized precision metrology \cite{suh2018soliton,yang2019vernier,trocha2018ultrafast} and coherent information processing \cite{marin2017microresonator,feldmann2021parallel}, CSs also serve as a canonical platform for exploring non-equilibrium dynamics \cite{cole2017soliton,weng2020heteronuclear}, offering an exceptional testbed for exploring complex topological phenomena \cite{mittal2021topological} and symmetry-breaking transitions in nonlinear driven systems \cite{garbin2020asymmetric,xu2021spontaneous}.

While CSs have predominantly been sustained by CW pumping, the extreme temporal mismatch between the highly localized CS and the uniform background imposes a fundamental bottleneck on energy efficiency \cite{bao2019laser}. Pulsed pumping architectures circumvent this limitation \cite{obrzud2017temporal}, but inherently couple the CS kinematics to the driving field gradient, a mechanism analogous to optical tweezing in the time domain \cite{jang2015temporal,hu2025dynamics,4hf5-r4nc}. Crucially, when the pump repetition rate is detuned from the cavity’s free spectral range (FSR), a regime known as desynchronous pumping, the interplay between Kerr nonlinearity and coherent driving forces CSs to deviate from the pump peak \cite{hendry2018spontaneous}. This temporal drift explicitly breaks the temporal reflection symmetry, compelling the CS to lock at an asymmetric, temporally offset position along the driving pulse flank, a state defined as temporal symmetry breaking (TSB).

Beyond these translational TSB dynamics, incorporating intrinsic vectorial degrees of freedom unlocks a fundamentally distinct landscape of internal symmetry breaking \cite{hu2025vector}. Physically, in coherently driven Kerr ring resonators, this internal asymmetry is driven by the intrinsic competition between intra-mode self-phase modulation (SPM) and inter-mode cross-phase modulation (XPM). When the XPM-induced nonlinear repulsion overcomes both the SPM confinement and the restoring force of the symmetric pump, the intracavity field spontaneously bifurcates into unequally populated polarization states. This spontaneous parity rupture, formally defined as polarization symmetry breaking (PSB), underpins the emergence of nontrivial macroscopic structures \cite{garbin2021dissipative,lucas2025polarization,nielsen2019coexistence}. Furthermore, this internal PSB is not an isolated phenomenon but rather a universal feature of dissipative systems across diverse photonic platforms \cite{cui2021xpm,yang2025polarization}. Harnessing this internal PSB has directly seeded advanced photonic technologies, such as coherent Ising machines \cite{quinn2026coherent}, ultrafast all-optical random number generators \cite{quinn2023random}, and highly sensitive polarization controllers \cite{moroney2022kerr}.

Historically, internal PSB and external TSB have been studied largely in isolation. While scalar desynchronous pumping merely displaces a single-component CS along the pulse flank without altering its internal symmetry \cite{hu2025dynamics,4hf5-r4nc,hendry2018spontaneous}, synchronously pulse-pumped vector resonators either pin solitons to the pump center \cite{hu2025vector} or rely on non-stationary parameter sweeps whose strong dependence on scanning rates obscures the deterministic steady-state topology \cite{copie2019interplay}. Furthermore, unlike uniform CW systems where directional and polarization degrees of freedom combine into additive multi-stage bifurcations \cite{hill2023multi}, coupling an internal \emph{spontaneous symmetry breaking}  to an external \emph{explicit symmetry breaking constraint} (asynchronous walk-off) pits two fundamentally distinct order parameters against each other. Whether a driven-dissipative system simply superimposes both asymmetries or resolves their competition through a non-trivial topological reorganization remains an open question critical to extending the Lugiato-Lefever paradigm.

In this Letter, we resolve this competition and demonstrate that tuning pump-cavity desynchronization induces a counterintuitive, symmetry-restoring phase transition rather than a trivial superposition of broken symmetries. Specifically, increasing desynchronization steers vector CSs from pure PSB, through a robust hybrid symmetry-breaking (HSB) regime where both asymmetries coexist, into a pure TSB state accompanied by the abrupt, spontaneous restoration of polarization parity. Using collective-coordinate variational perturbation theory, we map the effective Ginzburg-Landau potential landscape spanned by temporal offset and polarization asymmetry, revealing how kinematic drift weakens local driving until a saddle-node bifurcation eliminates the temporal confining barrier. Corroborated by iterated Ikeda-map simulations, these findings uncover a fundamental dynamic trade-off in which a driven-dissipative system sheds internal polarization complexity to accommodate external kinematic constraints.

\begin{figure*}[htb!]
    \centering
    \includegraphics[width=0.85\textwidth]{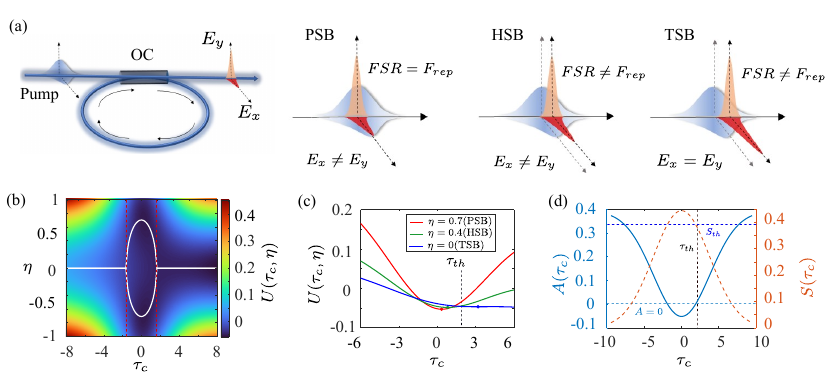}
    \caption{Competing symmetry breakings of cavity solitons (CSs) in a pulse-pumped fiber ring Kerr resonator. 
    (a) Schematic of the resonator supporting two orthogonal polarization modes ($E_x, E_y$), illustrating the three distinct dynamical regimes: pure polarization symmetry breaking (PSB: $E_x \neq E_y$, $\text{FSR} = F_{\text{rep}}$), hybrid symmetry breaking (HSB: $E_x \neq E_y$, $\text{FSR} \neq F_{\text{rep}}$), and pure temporal symmetry breaking (TSB: $E_x = E_y$, $\text{FSR} \neq F_{\text{rep}}$). OC: optical coupler; FSR: free spectral range; $F_{\text{rep}}$: pump repetition rate. 
    (b) 2D projection of the effective potential landscape $U(\tau_c, \eta)$ at a fixed asynchronous drift $v_d$. The white solid curve traces the equilibrium manifold of local potential minima, forming a closed pitchfork bifurcation bounded by the critical temporal thresholds $\pm\tau_{\text{th}}$ (vertical red dashed lines). 
    (c) Cross-sectional potential profiles along $\tau_c$ for $\eta = 0.7$ (PSB, red), $\eta = 0.4$ (HSB, green), and $\eta = 0$ (TSB, blue), with filled circles marking the respective potential minima, demonstrating the energy-minimum reversal from PSB near the pump center to TSB beyond $\tau_{\text{th}}$ (vertical black dashed line). 
    (d) Quadratic bifurcation coefficient $A(\tau_c)$ (solid blue line, left axis) and localized Gaussian pump profile $S(\tau_c)$ (dashed orange line, right axis). The zero-crossing of $A(\tau_c)$ ($A = 0$, horizontal dashed line) coincides with $S(\tau_c)$ dropping below the critical threshold $S_{\text{th}}$ (horizontal dotted line) at $\tau_c = \tau_{\text{th}}$. Dimensionless parameters: $d_2 = 1$, $\beta = 2.0$, $\Delta = 1.43$, $\sigma = 3.78$, $S_0 = 0.44$, $S_{\mathrm{th}} = 0.39$, and $\tau_{\mathrm{th}} = 1.89$ (see Supplemental Material for cavity parameters).}
    \label{fig1}
\end{figure*}

%%%%%%%%%%%%%%%%%%%%%%%%%%%%%%%%%%%%%%%%%
Figure~\ref{fig1}(a) illustrates a pulse-pumped passive Kerr fiber resonator supporting two orthogonal polarization modes $(E_x, E_y)$ via an optical coupler (OC), together with the three competing symmetry-breaking regimes. Under synchronous driving ($F_{\mathrm{rep}} = \mathrm{FSR}$), the soliton locks to the Gaussian pump peak while XPM induces spontaneous PSB ($E_x \neq E_y$). Introducing a repetition-rate mismatch ($F_{\mathrm{rep}} \neq \mathrm{FSR}$) displaces the soliton along the driving pulse flank into a HSB state where temporal offset and polarization imbalance ($E_x \neq E_y$) coexist, until stronger desynchronization pushes it further onto the pulse tail into a pure TSB state with restored polarization parity ($E_x = E_y$). Their mean-field evolution is governed by the normalized coupled Lugiato-Lefever equations (LLE):

\begin{align}
    \frac{\partial E_{x,y}}{\partial t} = & -(1+i\Delta)E_{x,y} + i\frac{d_2}{2}\frac{\partial^2 E_{x,y}}{\partial\tau^2}   \nonumber \\
    & - v_d\frac{\partial E_{x,y}}{\partial\tau} + iN_{x,y}E_{x,y} + S(\tau)
    \label{eq1}
\end{align}
\noindent where $t$ and $\tau$ are the dimensionless slow- and fast-time coordinates, respectively, $\Delta$ is the cavity detuning, and $d_2$ is the dispersion coefficient. The term $-v_d \partial_\tau E_{x,y}$ introduces a continuous temporal drift velocity $v_d$ arising from the asynchronous pumping mismatch. The nonlinear Kerr response $N_{x,y} = |E_{x,y}|^2 + \beta|E_{y,x}|^2$ couples the two modes via XPM of strength $\beta$. Both polarization modes are equally driven by a localized Gaussian pump $S(\tau) = S_0 \exp[-\tau^2/(2\sigma^2)]$ of peak amplitude $S_0$ and pulse width $\sigma$.

To analytically capture the competing symmetry-breaking dynamics in this driven-dissipative system, we project the coupled LLEs [Eq.~(\ref{eq1})] onto two slow collective coordinates via perturbation theory (see Supplemental Material): the soliton temporal centroid $\tau_c(t)$ relative to the pump center ($\tau=0$) and the fractional polarization asymmetry $\eta(t) \in [-1,1]$ between $E_x$ and $E_y$. Exploiting the mode-exchange symmetry ($E_x \leftrightarrow E_y$, i.e., $\eta \leftrightarrow -\eta$) and expanding the driven-dissipative overlap integrals in the weak-asymmetry limit yields a $\tau_c$-dependent Ginzburg-Landau effective potential:

\begin{equation}
    U(\tau_c, \eta) \approx U_0(\tau_c) - V_d \tau_c + A(\tau_c)\eta^2 + B(\tau_c)\eta^4, \tag{2}
\end{equation}
where $U_0(\tau_c) = -C_p S(\tau_c)$ ($C_p > 0$) is the $\eta$-independent temporal tweezing potential induced by the Gaussian pump envelope, $-V_d \tau_c$ ($V_d \propto v_d$) represents the linear tilt arising from asynchronous temporal walk-off, and $B(\tau_c) > 0$ guarantees global boundedness. Crucially, retaining the $\eta^0$ tweezing term $U_0(\tau_c)$, which is indispensable for spatially inhomogeneous driving, ensures a well-defined confining potential even when polarization parity is restored ($\eta = 0$).   The quadratic bifurcation coefficient $A(\tau_c)$ explicitly encodes the competition between linear cavity stabilization and pump-sustained inter-modal XPM repulsion:

\begin{equation}
    A(\tau_c) = \kappa_{\mathrm{lin}}\sqrt{d_2 \Delta} - \kappa_{\mathrm{NL}}(\beta - 1)\sqrt{d_2} S_0 \exp\!\left(-\frac{\tau_c^2}{2\sigma^2}\right) 
    \tag{3}
\end{equation}
where $\kappa_{\mathrm{lin}}, \kappa_{\mathrm{NL}} > 0$ are dimensionless constants set by the soliton profile integrals. Setting $A(\tau_c) = 0$ defines the critical local driving threshold for PSB:

\begin{equation}
   S_{\mathrm{th}} \equiv S_0 \exp\!\left(-\frac{\tau_c^2}{2\sigma^2}\right) = \frac{\kappa_{\mathrm{lin}}}{\kappa_{\mathrm{NL}}(\beta - 1)}\sqrt{\Delta}
   \tag{4} 
   \label{eq:4}
\end{equation}
which yields the critical temporal offset $\tau_{\mathrm{th}} = \sigma\sqrt{2\ln(S_0/S_{\mathrm{th}})}$. Inside $\vert{}\tau_c\vert{} < \tau_{\mathrm{th}}$, $A(\tau_c) < 0$ destabilizes the symmetric state ($\eta = 0$) into a double-well polarization potential ($\eta \neq 0$), whereas for $\vert{}\tau_c\vert{} > \tau_{\mathrm{th}}$, $A(\tau_c) > 0$ restores a single symmetric minimum at $\eta = 0$.   

By substituting realistic fiber-cavity parameters into the dimensionless LLE framework (see Supplemental Material), we map the effective potential $U(\tau_c, \eta)$ governing the steady-state vector CSs. While the potential adopts a symmetric double-well structure at $v_d = 0$ that locks the spontaneous PSB state at $\tau_c = 0$ ($\vert{}\eta\vert{} \approx 0.7$), introducing a finite asynchronous drift tilts the landscape along $\tau_c$, as shown by the 2D projection in Fig. \ref{fig1}(b) for a representative drift $v_d \approx 0.027$ (corresponding to $\kappa = 0.15\text{ ps/km}$). The white curve traces the equilibrium manifold of local potential minima, forming a closed pitchfork bifurcation bounded by the critical temporal thresholds $\pm\tau_{\mathrm{th}}$ (red dashed lines). Inside $\vert{}\tau_c\vert{} < \tau_{\mathrm{th}}$, spontaneous polarization breaking lowers the effective 1D potential by a condensation energy $-A(\tau_c)^2/[4B(\tau_c)]$, carving a deep, steep trapping well near the pump center on top of the broader Gaussian tweezing potential $U_0(\tau_c)$, whereas beyond $\vert{}\tau_c\vert{} > \tau_{\mathrm{th}}$ the two branches merge back to $\eta = 0$.   

Figure \ref{fig1}(c) compares the corresponding cross-sectional potential profiles along $\tau_c$ for $\eta = 0.7$ (PSB), $0.4$ (HSB), and $0$ (TSB), directly revealing an energy-minimum reversal. Near the pump center ($\tau_c \approx 0$), the PSB branch ($\eta = 0.7$) lies lowest in energy, whereas beyond $\tau_{\text{th}}$ the curves cross, and the TSB branch ($\eta = 0$) becomes the lowest-energy configuration. Meanwhile, the equilibrium minimum along each curve (filled circles) shifts progressively outward from $\tau_c \approx 0$ (PSB) to $\tau_c \approx 3.25$ (TSB). This topological transition is explicitly quantified in Fig. \ref{fig1}(d), where the bifurcation coefficient $A(\tau_c)$ crosses zero from negative to positive precisely as the Gaussian pump $S(\tau_c)$ falls below the critical threshold $S_{\mathrm{th}}$ at $\tau_c = \tau_{\mathrm{th}}$. Consequently, although the static equilibrium manifold is continuous in $\tau_c$, increasing $v_d$ beyond the maximum restoring gradient of the inner HSB well eliminates its confining barrier, triggering an abrupt jump into the outer pure TSB potential minimum with restored polarization parity. 

While this analytical framework elegantly unveils the topological origins of the symmetry breaking dynamics, it inherently relies on the mean-field approximation. Real-world pulsed-pumped resonators, however, often operate in regimes with localized coupling and non-negligible discrete round-trip variations that can disrupt mean-field dynamics. To rigorously validate our theoretical predictions beyond the mean-field limit, we numerically investigate the complete spatiotemporal evolution using an iterative round-trip Ikeda map (Supplementary Material). Crucially, the macroscopic drift $v_d$ from the analytical model is not phenomenologically imposed in the coupled nonlinear Schrödinger equations (NLSEs), instead, the asynchronous mismatch between the pump repetition rate and the cavity FSR is fundamentally incorporated through an advection term $-\kappa \partial E_{x,y} / \partial T$. This temporal walk-off parameter $\kappa$ serves as the microscopic physical counterpart to $v_d$, quantifying the temporal shift per unit propagation distance. By systematically tuning $\kappa$ in our full numerical simulations, we can effectively emulate the experimental mismatch between the pump repetition rate and the cavity FSR, thereby rigorously corroborating our analytical force-balance model and dynamically tracking the strictly ordered physical evolution of symmetry breaking.

\begin{figure}[htb]
	\centering
	\includegraphics[width=\linewidth]{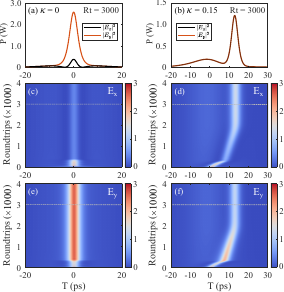}
    \caption{Evolution dynamics of the orthogonal polarization components $E_x$ and $E_y$ under different pump-cavity desynchronization levels $\kappa$. (a), (b) Steady-state intensity profiles of $\vert{}E_x\vert{}^2$ and $\vert{}E_y\vert{}^2$ at roundtrip 3000 for $\kappa = 0$ and $\kappa = 0.15$, respectively. (c), (e) Pseudocolor maps illustrating the temporal evolution of $E_x$ and $E_y$ over 4000 roundtrips at $\kappa = 0$. (d), (f) Corresponding temporal evolution of $E_x$ and $E_y$ at $\kappa = 0.15$. Parameters: $\delta_{x,y} = 0.2$, $P_0 = 0.09~\mathrm{W}$, $L = 100~\mathrm{m}$, $\beta_2 = -22~\mathrm{ps^2/km}$, $\gamma = 1.2~\mathrm{W^{-1}km^{-1}}$, $\theta = 0.05$, $\alpha = 0.28$, and $\tau_0 = 15~\mathrm{ps}$.}
    \label{fig2}
\end{figure}

Fig. \ref{fig2} presents the buildup of vector CSs at a fixed cavity detuning $\delta_{x,y} = 0.2$ and pump peak power $P_{0} = 0.09$ W, for two representative desynchronization conditions. In the synchronously pumped regime with $\kappa = 0$, the soliton stabilizes precisely at the temporal center of the pump pulse with zero net drift, as shown in the left column of Fig.~\ref{fig2}. A prominent power imbalance spontaneously arises between the two polarization components, confirming that PSB occurs even under perfectly symmetric driving, consistent with the theoretical predictions presented above. When a finite desynchronization $\kappa = 0.15$ ps/km is introduced, as shown in the right column of Fig.~\ref{fig2}, the soliton breaks its positional lock to the pump center and undergoes continuous temporal drift throughout the buildup process, marking the onset of TSB. Strikingly, the power disparity between $E_x$ and $E_y$ disappears concurrently in this drifting state. The two components recover nearly identical pulse intensities, demonstrating that polarization symmetry is spontaneously restored at the expense of translational symmetry. This direct numerical observation confirms the competitive nature of the two symmetry-breaking modes, showing that both the stable CS position ($T_c = 12.89$ ps, corresponding to $\tau_c \approx 3.25$ in Fig. \ref{fig1}(c)) and the resulting symmetry-breaking states are in excellent agreement with preceding analytical predictions. The system relinquishes its internal polarization asymmetry to accommodate the external drift imposed by desynchronized pumping.

\begin{figure}[htb!]
	\centering
	\includegraphics[width=\linewidth]{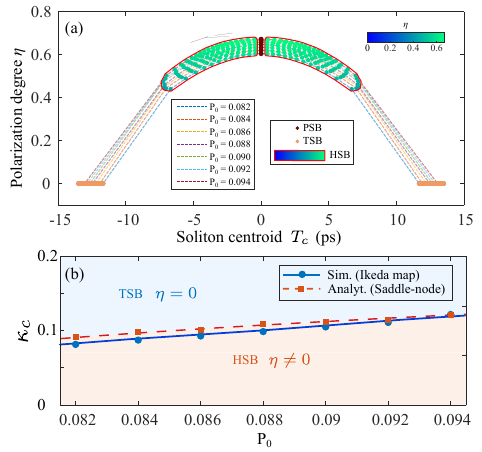}
    \caption{Phase diagram of the competing symmetry-breaking dynamics. (a) Two-dimensional phase space spanned by the soliton temporal centroid $T_c$ and polarization asymmetry $\eta$ (showing the $\eta \ge 0$ branch by mode-exchange symmetry), corroborating the upper equilibrium manifold in Fig.~\ref{fig1}(b). Trajectories (dashed lines) are obtained by scanning the desynchronization $\kappa \in [-0.2, 0.2]~\mathrm{ps/km}$ for pump peak powers $P_0$ from $0.082$ to $0.094~\mathrm{W}$ in steps of $0.002~\mathrm{W}$. Markers denote pure PSB (dark-red circles at $T_c = 0$), pure TSB (orange circles at $\eta = 0$), and the intermediate HSB (blue-green gradient enclosed by the solid red boundary). (b) Critical desynchronization threshold $\kappa_c$ for the abrupt HSB-to-TSB transition versus pump peak power $P_0$, comparing full Ikeda-map simulations (solid blue line with circles) with the analytical saddle-node phase boundary (dashed red line with squares). Shaded regions indicate the HSB ($\eta \neq 0$) and pure TSB ($\eta = 0$) phases. Fixed cavity detuning: $\delta_{x,y} = 0.2$; other parameters are the same as in Fig.~\ref{fig2}.}
  \label{fig3}
\end{figure}

To directly test the effective potential topology of Fig.~\ref{fig1} across a broader parameter space, we map the phase diagram in the $(T_c, \eta)$ plane by scanning $\kappa \in [-0.2, 0.2]~\text{ps/km}$ for pump peak powers $P_0 \in [82, 94]~\text{mW}$ [Fig.~\ref{fig3}(a)]. Remarkably, the numerical trajectories directly reproduce the upper half of the theoretical pitchfork manifold predicted in Fig.~\ref{fig1}(b). As $|\kappa|$ increases, the soliton evolves continuously from the central pure PSB state ($\tau_c = 0$, red circles) along the arched HSB branch (blue-green gradient). Crucially, rather than following the static manifold smoothly down to $\eta = 0$, each trajectory terminates near $|T_c| \approx 7~\text{ps}$ ($\eta \approx 0.45$) and jumps abruptly (dashed lines) to the outer pure TSB branch ($|T_c| \approx 12.5~\text{ps}$, $\eta = 0$, orange circles). This discontinuous leap is the direct dynamical signature of the saddle-node barrier collapse revealed in Fig.~\ref{fig1}(c). Furthermore, Fig.~\ref{fig3}(b) quantitatively validates the analytical phase boundary derived by combining the threshold condition [Eq.~(\ref{eq:4})] with the force-balance relation, which predicts $\kappa_c(P_0) \propto \sqrt{\ln(P_0/P_{0,\text{th}})}$ (see Supplemental Material). Across the entire power range, the numerically extracted critical thresholds (solid blue) closely follow the analytical phase boundary (dashed red), with the slight upward offset of the analytical curve arising because $\kappa_c$ evaluates the force balance at the bifurcation boundary $\tau_{\mathrm{th}}$ rather than the slightly inner saddle-node point ($\tau_{\mathrm{SN}} < \tau_{\mathrm{th}}$), alongside finite-loss effects beyond the mean-field limit.

\begin{figure}[t!]
	\centering
	\includegraphics[width=\linewidth]{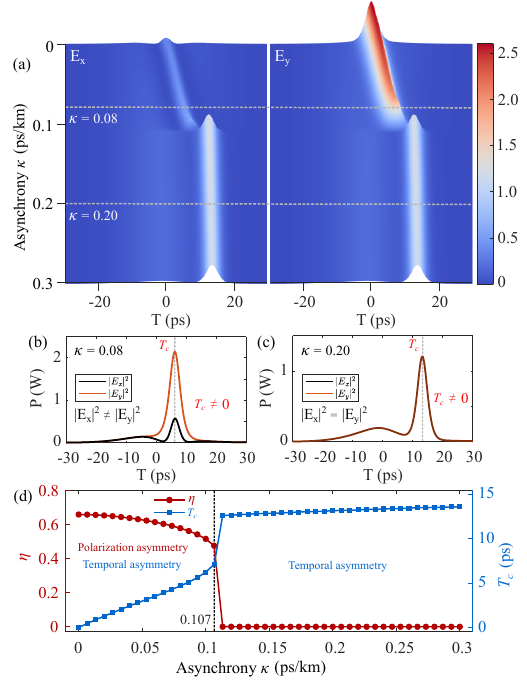}
     \caption{Dynamics of the PSB-HSB-TSB phase transition. (a) Steady-state temporal evolution of the vector soliton intensities ($E_x$, left; $E_y$, right) under a continuous scan of $\kappa$. (b),(c) Representative intensity profiles at $\kappa = 0.08$ and $0.20$ ps/km [gray dashed lines in (a)], explicitly illustrating the intermediate HSB and pure TSB states, respectively. Vertical dashed lines mark the temporal peak position $T_c$. (d) Bifurcation diagram detailing the evolution of the polarization asymmetry $\eta$ (red circles) and $T_c$ (blue squares) versus $\kappa$, highlighting the abrupt phase transition at the critical threshold. Parameters: $\delta_{x,y} = 0.2$ and $P_0 = 0.09~\mathrm{W}$ ($90~\mathrm{mW}$), other parameters are the same as in Fig.~2.}
     \label{fig4}
\end{figure}

To quantitatively characterize these symmetry-breaking dynamics, Fig. \ref{fig4}(a) presents the simulation results of the steady-state temporal intensity evolution of $E_x$ (left) and $E_y$ (right) as a function of $\kappa$. Representative intensity profiles at $\kappa = \SI{0.08}{ps/km}$ and $\SI{0.20}{ps/km}$ (indicated by gray dashed lines in Fig. \ref{fig4}(a) are detailed in Fig. \ref{fig4}(b) and (c), visually illustrating the HSB and pure TSB states, respectively. The corresponding bifurcation diagram in Fig. \ref{fig4}(d) summarizes the dependence of the polarization asymmetry $\eta$ and the soliton centroid position $T_c$ on $\kappa$, revealing two distinct regimes separated by a critical threshold $\kappa_c = \SI{0.107}{ps/km}$. Below $\kappa_c$, PSB and TSB coexist, with the temporal offset becoming progressively more pronounced as $\kappa$ increases. Above $\kappa_c$, the polarization asymmetry abruptly vanishes ($\eta=0$), and the system transitions into a pure TSB state. This abrupt phase transition exposes a fundamental dynamic compromise, in which the system sacrifices internal polarization complexity to accommodate the external kinematic drift, thereby spontaneously restoring polarization parity.

In conclusion, we have demonstrated that tuning pump-cavity desynchronization in driven Kerr resonators governs the dynamic competition between internal spontaneous PSB and external explicit TSB, steering vector CSs from pure PSB, through a robust intermediate HSB regime, into pure TSB accompanied by the abrupt restoration of polarization parity. Combining collective-coordinate perturbation analysis of the coupled LLEs with full iterated Ikeda-map simulations, we established how drift-induced temporal displacement weakens local driving until the temporal confining barrier vanishes, forcing the system to shed internal polarization complexity to satisfy external kinematic constraints. This mechanism extends the Lugiato-Lefever paradigm into coupled vector-field dynamics, providing a controllable photonic analog to competing-order-parameter and multiferroic phase transitions in condensed matter. Because these regimes are accessible under realistic standard-fiber cavity parameters, macroscopic desynchronization offers a direct, reconfigurable experimental knob for navigating multidimensional symmetry-breaking landscapes, with promising implications for all-optical switching and multiplexed photonic information processing.

%%%%%%%%%%%%%%%%%%%%%%%%%%%%%%%%%%%%%%%%%%%%
\section*{ACKNOWLEDGMENTS}
This work was supported by the National Natural Science Foundation of China (No.62375001, 62175116), the Talent start-up fund of Anhui University (No. S020318030/009), Overseas talent project of Ministry of Education (No. S202518001/002), and Anhui Provincial Department of Education Outstanding Scientific Research and Innovation Team Project (2025AHGXZK10012).

\bibliography{library}

\begin{thebibliography}{32}%
\makeatletter
\providecommand \@ifxundefined [1]{%
 \@ifx{#1\undefined}
}%
\providecommand \@ifnum [1]{%
 \ifnum #1\expandafter \@firstoftwo
 \else \expandafter \@secondoftwo
 \fi
}%
\providecommand \@ifx [1]{%
 \ifx #1\expandafter \@firstoftwo
 \else \expandafter \@secondoftwo
 \fi
}%
\providecommand \natexlab [1]{#1}%
\providecommand \enquote  [1]{``#1''}%
\providecommand \bibnamefont  [1]{#1}%
\providecommand \bibfnamefont [1]{#1}%
\providecommand \citenamefont [1]{#1}%
\providecommand \href@noop [0]{\@secondoftwo}%
\providecommand \href [0]{\begingroup \@sanitize@url \@href}%
\providecommand \@href[1]{\@@startlink{#1}\@@href}%
\providecommand \@@href[1]{\endgroup#1\@@endlink}%
\providecommand \@sanitize@url [0]{\catcode `\\12\catcode `\$12\catcode
  `\&12\catcode `\#12\catcode `\^12\catcode `\_12\catcode `\%12\relax}%
\providecommand \@@startlink[1]{}%
\providecommand \@@endlink[0]{}%
\providecommand \url  [0]{\begingroup\@sanitize@url \@url }%
\providecommand \@url [1]{\endgroup\@href {#1}{\urlprefix }}%
\providecommand \urlprefix  [0]{URL }%
\providecommand \Eprint [0]{\href }%
\providecommand \doibase [0]{https://doi.org/}%
\providecommand \selectlanguage [0]{\@gobble}%
\providecommand \bibinfo  [0]{\@secondoftwo}%
\providecommand \bibfield  [0]{\@secondoftwo}%
\providecommand \translation [1]{[#1]}%
\providecommand \BibitemOpen [0]{}%
\providecommand \bibitemStop [0]{}%
\providecommand \bibitemNoStop [0]{.\EOS\space}%
\providecommand \EOS [0]{\spacefactor3000\relax}%
\providecommand \BibitemShut  [1]{\csname bibitem#1\endcsname}%
\let\auto@bib@innerbib\@empty
%</preamble>
\bibitem [{\citenamefont {Cross}\ and\ \citenamefont
  {Hohenberg}(1993)}]{cross1993pattern}%
  \BibitemOpen
  \bibfield  {author} {\bibinfo {author} {\bibfnamefont {M.~C.}\ \bibnamefont
  {Cross}}\ and\ \bibinfo {author} {\bibfnamefont {P.~C.}\ \bibnamefont
  {Hohenberg}},\ }\bibfield  {title} {\bibinfo {title} {Pattern formation
  outside of equilibrium},\ }\href@noop {} {\bibfield  {journal} {\bibinfo
  {journal} {Reviews of modern physics}\ }\textbf {\bibinfo {volume} {65}},\
  \bibinfo {pages} {851} (\bibinfo {year} {1993})}\BibitemShut {NoStop}%
\bibitem [{\citenamefont {Grelu}\ and\ \citenamefont
  {Akhmediev}(2012)}]{grelu2012dissipative}%
  \BibitemOpen
  \bibfield  {author} {\bibinfo {author} {\bibfnamefont {P.}~\bibnamefont
  {Grelu}}\ and\ \bibinfo {author} {\bibfnamefont {N.}~\bibnamefont
  {Akhmediev}},\ }\bibfield  {title} {\bibinfo {title} {Dissipative solitons
  for mode-locked lasers},\ }\href@noop {} {\bibfield  {journal} {\bibinfo
  {journal} {Nature photonics}\ }\textbf {\bibinfo {volume} {6}},\ \bibinfo
  {pages} {84} (\bibinfo {year} {2012})}\BibitemShut {NoStop}%
\bibitem [{\citenamefont {Leo}\ \emph {et~al.}(2010)\citenamefont {Leo},
  \citenamefont {Coen}, \citenamefont {Kockaert}, \citenamefont {Gorza},
  \citenamefont {Emplit},\ and\ \citenamefont {Haelterman}}]{leo2010temporal}%
  \BibitemOpen
  \bibfield  {author} {\bibinfo {author} {\bibfnamefont {F.}~\bibnamefont
  {Leo}}, \bibinfo {author} {\bibfnamefont {S.}~\bibnamefont {Coen}}, \bibinfo
  {author} {\bibfnamefont {P.}~\bibnamefont {Kockaert}}, \bibinfo {author}
  {\bibfnamefont {S.-P.}\ \bibnamefont {Gorza}}, \bibinfo {author}
  {\bibfnamefont {P.}~\bibnamefont {Emplit}},\ and\ \bibinfo {author}
  {\bibfnamefont {M.}~\bibnamefont {Haelterman}},\ }\bibfield  {title}
  {\bibinfo {title} {Temporal cavity solitons in one-dimensional kerr media as
  bits in an all-optical buffer},\ }\href@noop {} {\bibfield  {journal}
  {\bibinfo  {journal} {Nature Photonics}\ }\textbf {\bibinfo {volume} {4}},\
  \bibinfo {pages} {471} (\bibinfo {year} {2010})}\BibitemShut {NoStop}%
\bibitem [{\citenamefont {Herr}\ \emph {et~al.}(2014)\citenamefont {Herr},
  \citenamefont {Brasch}, \citenamefont {Jost}, \citenamefont {Wang},
  \citenamefont {Kondratiev}, \citenamefont {Gorodetsky},\ and\ \citenamefont
  {Kippenberg}}]{herr2014temporal}%
  \BibitemOpen
  \bibfield  {author} {\bibinfo {author} {\bibfnamefont {T.}~\bibnamefont
  {Herr}}, \bibinfo {author} {\bibfnamefont {V.}~\bibnamefont {Brasch}},
  \bibinfo {author} {\bibfnamefont {J.~D.}\ \bibnamefont {Jost}}, \bibinfo
  {author} {\bibfnamefont {C.~Y.}\ \bibnamefont {Wang}}, \bibinfo {author}
  {\bibfnamefont {N.~M.}\ \bibnamefont {Kondratiev}}, \bibinfo {author}
  {\bibfnamefont {M.~L.}\ \bibnamefont {Gorodetsky}},\ and\ \bibinfo {author}
  {\bibfnamefont {T.~J.}\ \bibnamefont {Kippenberg}},\ }\bibfield  {title}
  {\bibinfo {title} {Temporal solitons in optical microresonators},\
  }\href@noop {} {\bibfield  {journal} {\bibinfo  {journal} {Nature Photonics}\
  }\textbf {\bibinfo {volume} {8}},\ \bibinfo {pages} {145} (\bibinfo {year}
  {2014})}\BibitemShut {NoStop}%
\bibitem [{\citenamefont {Lilienfein}\ \emph {et~al.}(2019)\citenamefont
  {Lilienfein}, \citenamefont {Hofer}, \citenamefont {H{\"o}gner},
  \citenamefont {Saule}, \citenamefont {Trubetskov}, \citenamefont {Pervak},
  \citenamefont {Fill}, \citenamefont {Riek}, \citenamefont {Leitenstorfer},
  \citenamefont {Limpert} \emph {et~al.}}]{lilienfein2019temporal}%
  \BibitemOpen
  \bibfield  {author} {\bibinfo {author} {\bibfnamefont {N.}~\bibnamefont
  {Lilienfein}}, \bibinfo {author} {\bibfnamefont {C.}~\bibnamefont {Hofer}},
  \bibinfo {author} {\bibfnamefont {M.}~\bibnamefont {H{\"o}gner}}, \bibinfo
  {author} {\bibfnamefont {T.}~\bibnamefont {Saule}}, \bibinfo {author}
  {\bibfnamefont {M.}~\bibnamefont {Trubetskov}}, \bibinfo {author}
  {\bibfnamefont {V.}~\bibnamefont {Pervak}}, \bibinfo {author} {\bibfnamefont
  {E.}~\bibnamefont {Fill}}, \bibinfo {author} {\bibfnamefont {C.}~\bibnamefont
  {Riek}}, \bibinfo {author} {\bibfnamefont {A.}~\bibnamefont {Leitenstorfer}},
  \bibinfo {author} {\bibfnamefont {J.}~\bibnamefont {Limpert}}, \emph
  {et~al.},\ }\bibfield  {title} {\bibinfo {title} {Temporal solitons in
  free-space femtosecond enhancement cavities},\ }\href@noop {} {\bibfield
  {journal} {\bibinfo  {journal} {Nature Photonics}\ }\textbf {\bibinfo
  {volume} {13}},\ \bibinfo {pages} {214} (\bibinfo {year} {2019})}\BibitemShut
  {NoStop}%
\bibitem [{\citenamefont {Suh}\ and\ \citenamefont
  {Vahala}(2018)}]{suh2018soliton}%
  \BibitemOpen
  \bibfield  {author} {\bibinfo {author} {\bibfnamefont {M.-G.}\ \bibnamefont
  {Suh}}\ and\ \bibinfo {author} {\bibfnamefont {K.~J.}\ \bibnamefont
  {Vahala}},\ }\bibfield  {title} {\bibinfo {title} {Soliton microcomb range
  measurement},\ }\href@noop {} {\bibfield  {journal} {\bibinfo  {journal}
  {Science}\ }\textbf {\bibinfo {volume} {359}},\ \bibinfo {pages} {884}
  (\bibinfo {year} {2018})}\BibitemShut {NoStop}%
\bibitem [{\citenamefont {Yang}\ \emph {et~al.}(2019)\citenamefont {Yang},
  \citenamefont {Shen}, \citenamefont {Wang}, \citenamefont {Tran},
  \citenamefont {Zhang}, \citenamefont {Yang}, \citenamefont {Wu},
  \citenamefont {Bao}, \citenamefont {Bowers}, \citenamefont {Yariv} \emph
  {et~al.}}]{yang2019vernier}%
  \BibitemOpen
  \bibfield  {author} {\bibinfo {author} {\bibfnamefont {Q.-F.}\ \bibnamefont
  {Yang}}, \bibinfo {author} {\bibfnamefont {B.}~\bibnamefont {Shen}}, \bibinfo
  {author} {\bibfnamefont {H.}~\bibnamefont {Wang}}, \bibinfo {author}
  {\bibfnamefont {M.}~\bibnamefont {Tran}}, \bibinfo {author} {\bibfnamefont
  {Z.}~\bibnamefont {Zhang}}, \bibinfo {author} {\bibfnamefont {K.~Y.}\
  \bibnamefont {Yang}}, \bibinfo {author} {\bibfnamefont {L.}~\bibnamefont
  {Wu}}, \bibinfo {author} {\bibfnamefont {C.}~\bibnamefont {Bao}}, \bibinfo
  {author} {\bibfnamefont {J.}~\bibnamefont {Bowers}}, \bibinfo {author}
  {\bibfnamefont {A.}~\bibnamefont {Yariv}}, \emph {et~al.},\ }\bibfield
  {title} {\bibinfo {title} {Vernier spectrometer using counterpropagating
  soliton microcombs},\ }\href@noop {} {\bibfield  {journal} {\bibinfo
  {journal} {Science}\ }\textbf {\bibinfo {volume} {363}},\ \bibinfo {pages}
  {965} (\bibinfo {year} {2019})}\BibitemShut {NoStop}%
\bibitem [{\citenamefont {Trocha}\ \emph {et~al.}(2018)\citenamefont {Trocha},
  \citenamefont {Karpov}, \citenamefont {Ganin}, \citenamefont {Pfeiffer},
  \citenamefont {Kordts}, \citenamefont {Wolf}, \citenamefont {Krockenberger},
  \citenamefont {Marin-Palomo}, \citenamefont {Weimann}, \citenamefont {Randel}
  \emph {et~al.}}]{trocha2018ultrafast}%
  \BibitemOpen
  \bibfield  {author} {\bibinfo {author} {\bibfnamefont {P.}~\bibnamefont
  {Trocha}}, \bibinfo {author} {\bibfnamefont {M.}~\bibnamefont {Karpov}},
  \bibinfo {author} {\bibfnamefont {D.}~\bibnamefont {Ganin}}, \bibinfo
  {author} {\bibfnamefont {M.~H.}\ \bibnamefont {Pfeiffer}}, \bibinfo {author}
  {\bibfnamefont {A.}~\bibnamefont {Kordts}}, \bibinfo {author} {\bibfnamefont
  {S.}~\bibnamefont {Wolf}}, \bibinfo {author} {\bibfnamefont {J.}~\bibnamefont
  {Krockenberger}}, \bibinfo {author} {\bibfnamefont {P.}~\bibnamefont
  {Marin-Palomo}}, \bibinfo {author} {\bibfnamefont {C.}~\bibnamefont
  {Weimann}}, \bibinfo {author} {\bibfnamefont {S.}~\bibnamefont {Randel}},
  \emph {et~al.},\ }\bibfield  {title} {\bibinfo {title} {Ultrafast optical
  ranging using microresonator soliton frequency combs},\ }\href@noop {}
  {\bibfield  {journal} {\bibinfo  {journal} {Science}\ }\textbf {\bibinfo
  {volume} {359}},\ \bibinfo {pages} {887} (\bibinfo {year}
  {2018})}\BibitemShut {NoStop}%
\bibitem [{\citenamefont {Marin-Palomo}\ \emph {et~al.}(2017)\citenamefont
  {Marin-Palomo}, \citenamefont {Kemal}, \citenamefont {Karpov}, \citenamefont
  {Kordts}, \citenamefont {Pfeifle}, \citenamefont {Pfeiffer}, \citenamefont
  {Trocha}, \citenamefont {Wolf}, \citenamefont {Brasch}, \citenamefont
  {Anderson} \emph {et~al.}}]{marin2017microresonator}%
  \BibitemOpen
  \bibfield  {author} {\bibinfo {author} {\bibfnamefont {P.}~\bibnamefont
  {Marin-Palomo}}, \bibinfo {author} {\bibfnamefont {J.~N.}\ \bibnamefont
  {Kemal}}, \bibinfo {author} {\bibfnamefont {M.}~\bibnamefont {Karpov}},
  \bibinfo {author} {\bibfnamefont {A.}~\bibnamefont {Kordts}}, \bibinfo
  {author} {\bibfnamefont {J.}~\bibnamefont {Pfeifle}}, \bibinfo {author}
  {\bibfnamefont {M.~H.}\ \bibnamefont {Pfeiffer}}, \bibinfo {author}
  {\bibfnamefont {P.}~\bibnamefont {Trocha}}, \bibinfo {author} {\bibfnamefont
  {S.}~\bibnamefont {Wolf}}, \bibinfo {author} {\bibfnamefont {V.}~\bibnamefont
  {Brasch}}, \bibinfo {author} {\bibfnamefont {M.~H.}\ \bibnamefont
  {Anderson}}, \emph {et~al.},\ }\bibfield  {title} {\bibinfo {title}
  {Microresonator-based solitons for massively parallel coherent optical
  communications},\ }\href@noop {} {\bibfield  {journal} {\bibinfo  {journal}
  {Nature}\ }\textbf {\bibinfo {volume} {546}},\ \bibinfo {pages} {274}
  (\bibinfo {year} {2017})}\BibitemShut {NoStop}%
\bibitem [{\citenamefont {Feldmann}\ \emph {et~al.}(2021)\citenamefont
  {Feldmann}, \citenamefont {Youngblood}, \citenamefont {Karpov}, \citenamefont
  {Gehring}, \citenamefont {Li}, \citenamefont {Stappers}, \citenamefont
  {Le~Gallo}, \citenamefont {Fu}, \citenamefont {Lukashchuk}, \citenamefont
  {Raja} \emph {et~al.}}]{feldmann2021parallel}%
  \BibitemOpen
  \bibfield  {author} {\bibinfo {author} {\bibfnamefont {J.}~\bibnamefont
  {Feldmann}}, \bibinfo {author} {\bibfnamefont {N.}~\bibnamefont
  {Youngblood}}, \bibinfo {author} {\bibfnamefont {M.}~\bibnamefont {Karpov}},
  \bibinfo {author} {\bibfnamefont {H.}~\bibnamefont {Gehring}}, \bibinfo
  {author} {\bibfnamefont {X.}~\bibnamefont {Li}}, \bibinfo {author}
  {\bibfnamefont {M.}~\bibnamefont {Stappers}}, \bibinfo {author}
  {\bibfnamefont {M.}~\bibnamefont {Le~Gallo}}, \bibinfo {author}
  {\bibfnamefont {X.}~\bibnamefont {Fu}}, \bibinfo {author} {\bibfnamefont
  {A.}~\bibnamefont {Lukashchuk}}, \bibinfo {author} {\bibfnamefont {A.~S.}\
  \bibnamefont {Raja}}, \emph {et~al.},\ }\bibfield  {title} {\bibinfo {title}
  {Parallel convolutional processing using an integrated photonic tensor
  core},\ }\href@noop {} {\bibfield  {journal} {\bibinfo  {journal} {Nature}\
  }\textbf {\bibinfo {volume} {589}},\ \bibinfo {pages} {52} (\bibinfo {year}
  {2021})}\BibitemShut {NoStop}%
\bibitem [{\citenamefont {Cole}\ \emph {et~al.}(2017)\citenamefont {Cole},
  \citenamefont {Lamb}, \citenamefont {Del’Haye}, \citenamefont {Diddams},\
  and\ \citenamefont {Papp}}]{cole2017soliton}%
  \BibitemOpen
  \bibfield  {author} {\bibinfo {author} {\bibfnamefont {D.~C.}\ \bibnamefont
  {Cole}}, \bibinfo {author} {\bibfnamefont {E.~S.}\ \bibnamefont {Lamb}},
  \bibinfo {author} {\bibfnamefont {P.}~\bibnamefont {Del’Haye}}, \bibinfo
  {author} {\bibfnamefont {S.~A.}\ \bibnamefont {Diddams}},\ and\ \bibinfo
  {author} {\bibfnamefont {S.~B.}\ \bibnamefont {Papp}},\ }\bibfield  {title}
  {\bibinfo {title} {Soliton crystals in kerr resonators},\ }\href@noop {}
  {\bibfield  {journal} {\bibinfo  {journal} {Nature Photonics}\ }\textbf
  {\bibinfo {volume} {11}},\ \bibinfo {pages} {671} (\bibinfo {year}
  {2017})}\BibitemShut {NoStop}%
\bibitem [{\citenamefont {Weng}\ \emph {et~al.}(2020)\citenamefont {Weng},
  \citenamefont {Bouchand}, \citenamefont {Lucas}, \citenamefont {Obrzud},
  \citenamefont {Herr},\ and\ \citenamefont
  {Kippenberg}}]{weng2020heteronuclear}%
  \BibitemOpen
  \bibfield  {author} {\bibinfo {author} {\bibfnamefont {W.}~\bibnamefont
  {Weng}}, \bibinfo {author} {\bibfnamefont {R.}~\bibnamefont {Bouchand}},
  \bibinfo {author} {\bibfnamefont {E.}~\bibnamefont {Lucas}}, \bibinfo
  {author} {\bibfnamefont {E.}~\bibnamefont {Obrzud}}, \bibinfo {author}
  {\bibfnamefont {T.}~\bibnamefont {Herr}},\ and\ \bibinfo {author}
  {\bibfnamefont {T.~J.}\ \bibnamefont {Kippenberg}},\ }\bibfield  {title}
  {\bibinfo {title} {Heteronuclear soliton molecules in optical
  microresonators},\ }\href@noop {} {\bibfield  {journal} {\bibinfo  {journal}
  {Nature communications}\ }\textbf {\bibinfo {volume} {11}},\ \bibinfo {pages}
  {2402} (\bibinfo {year} {2020})}\BibitemShut {NoStop}%
\bibitem [{\citenamefont {Mittal}\ \emph {et~al.}(2021)\citenamefont {Mittal},
  \citenamefont {Moille}, \citenamefont {Srinivasan}, \citenamefont {Chembo},\
  and\ \citenamefont {Hafezi}}]{mittal2021topological}%
  \BibitemOpen
  \bibfield  {author} {\bibinfo {author} {\bibfnamefont {S.}~\bibnamefont
  {Mittal}}, \bibinfo {author} {\bibfnamefont {G.}~\bibnamefont {Moille}},
  \bibinfo {author} {\bibfnamefont {K.}~\bibnamefont {Srinivasan}}, \bibinfo
  {author} {\bibfnamefont {Y.~K.}\ \bibnamefont {Chembo}},\ and\ \bibinfo
  {author} {\bibfnamefont {M.}~\bibnamefont {Hafezi}},\ }\bibfield  {title}
  {\bibinfo {title} {Topological frequency combs and nested temporal
  solitons},\ }\href@noop {} {\bibfield  {journal} {\bibinfo  {journal} {Nature
  Physics}\ }\textbf {\bibinfo {volume} {17}},\ \bibinfo {pages} {1169}
  (\bibinfo {year} {2021})}\BibitemShut {NoStop}%
\bibitem [{\citenamefont {Garbin}\ \emph {et~al.}(2020)\citenamefont {Garbin},
  \citenamefont {Fatome}, \citenamefont {Oppo}, \citenamefont {Erkintalo},
  \citenamefont {Murdoch},\ and\ \citenamefont {Coen}}]{garbin2020asymmetric}%
  \BibitemOpen
  \bibfield  {author} {\bibinfo {author} {\bibfnamefont {B.}~\bibnamefont
  {Garbin}}, \bibinfo {author} {\bibfnamefont {J.}~\bibnamefont {Fatome}},
  \bibinfo {author} {\bibfnamefont {G.-L.}\ \bibnamefont {Oppo}}, \bibinfo
  {author} {\bibfnamefont {M.}~\bibnamefont {Erkintalo}}, \bibinfo {author}
  {\bibfnamefont {S.~G.}\ \bibnamefont {Murdoch}},\ and\ \bibinfo {author}
  {\bibfnamefont {S.}~\bibnamefont {Coen}},\ }\bibfield  {title} {\bibinfo
  {title} {Asymmetric balance in symmetry breaking},\ }\href@noop {} {\bibfield
   {journal} {\bibinfo  {journal} {Physical Review Research}\ }\textbf
  {\bibinfo {volume} {2}},\ \bibinfo {pages} {023244} (\bibinfo {year}
  {2020})}\BibitemShut {NoStop}%
\bibitem [{\citenamefont {Xu}\ \emph {et~al.}(2021)\citenamefont {Xu},
  \citenamefont {Nielsen}, \citenamefont {Garbin}, \citenamefont {Hill},
  \citenamefont {Oppo}, \citenamefont {Fatome}, \citenamefont {Murdoch},
  \citenamefont {Coen},\ and\ \citenamefont {Erkintalo}}]{xu2021spontaneous}%
  \BibitemOpen
  \bibfield  {author} {\bibinfo {author} {\bibfnamefont {G.}~\bibnamefont
  {Xu}}, \bibinfo {author} {\bibfnamefont {A.~U.}\ \bibnamefont {Nielsen}},
  \bibinfo {author} {\bibfnamefont {B.}~\bibnamefont {Garbin}}, \bibinfo
  {author} {\bibfnamefont {L.}~\bibnamefont {Hill}}, \bibinfo {author}
  {\bibfnamefont {G.-L.}\ \bibnamefont {Oppo}}, \bibinfo {author}
  {\bibfnamefont {J.}~\bibnamefont {Fatome}}, \bibinfo {author} {\bibfnamefont
  {S.~G.}\ \bibnamefont {Murdoch}}, \bibinfo {author} {\bibfnamefont
  {S.}~\bibnamefont {Coen}},\ and\ \bibinfo {author} {\bibfnamefont
  {M.}~\bibnamefont {Erkintalo}},\ }\bibfield  {title} {\bibinfo {title}
  {Spontaneous symmetry breaking of dissipative optical solitons in a
  two-component kerr resonator},\ }\href@noop {} {\bibfield  {journal}
  {\bibinfo  {journal} {Nature Communications}\ }\textbf {\bibinfo {volume}
  {12}},\ \bibinfo {pages} {4023} (\bibinfo {year} {2021})}\BibitemShut
  {NoStop}%
\bibitem [{\citenamefont {Bao}\ \emph {et~al.}(2019)\citenamefont {Bao},
  \citenamefont {Cooper}, \citenamefont {Rowley}, \citenamefont {Di~Lauro},
  \citenamefont {Totero~Gongora}, \citenamefont {Chu}, \citenamefont {Little},
  \citenamefont {Oppo}, \citenamefont {Morandotti}, \citenamefont {Moss} \emph
  {et~al.}}]{bao2019laser}%
  \BibitemOpen
  \bibfield  {author} {\bibinfo {author} {\bibfnamefont {H.}~\bibnamefont
  {Bao}}, \bibinfo {author} {\bibfnamefont {A.}~\bibnamefont {Cooper}},
  \bibinfo {author} {\bibfnamefont {M.}~\bibnamefont {Rowley}}, \bibinfo
  {author} {\bibfnamefont {L.}~\bibnamefont {Di~Lauro}}, \bibinfo {author}
  {\bibfnamefont {J.~S.}\ \bibnamefont {Totero~Gongora}}, \bibinfo {author}
  {\bibfnamefont {S.~T.}\ \bibnamefont {Chu}}, \bibinfo {author} {\bibfnamefont
  {B.~E.}\ \bibnamefont {Little}}, \bibinfo {author} {\bibfnamefont {G.-L.}\
  \bibnamefont {Oppo}}, \bibinfo {author} {\bibfnamefont {R.}~\bibnamefont
  {Morandotti}}, \bibinfo {author} {\bibfnamefont {D.~J.}\ \bibnamefont
  {Moss}}, \emph {et~al.},\ }\bibfield  {title} {\bibinfo {title} {Laser
  cavity-soliton microcombs},\ }\href@noop {} {\bibfield  {journal} {\bibinfo
  {journal} {Nature Photonics}\ }\textbf {\bibinfo {volume} {13}},\ \bibinfo
  {pages} {384} (\bibinfo {year} {2019})}\BibitemShut {NoStop}%
\bibitem [{\citenamefont {Obrzud}\ \emph {et~al.}(2017)\citenamefont {Obrzud},
  \citenamefont {Lecomte},\ and\ \citenamefont {Herr}}]{obrzud2017temporal}%
  \BibitemOpen
  \bibfield  {author} {\bibinfo {author} {\bibfnamefont {E.}~\bibnamefont
  {Obrzud}}, \bibinfo {author} {\bibfnamefont {S.}~\bibnamefont {Lecomte}},\
  and\ \bibinfo {author} {\bibfnamefont {T.}~\bibnamefont {Herr}},\ }\bibfield
  {title} {\bibinfo {title} {Temporal solitons in microresonators driven by
  optical pulses},\ }\href@noop {} {\bibfield  {journal} {\bibinfo  {journal}
  {Nature Photonics}\ }\textbf {\bibinfo {volume} {11}},\ \bibinfo {pages}
  {600} (\bibinfo {year} {2017})}\BibitemShut {NoStop}%
\bibitem [{\citenamefont {Jang}\ \emph {et~al.}(2015)\citenamefont {Jang},
  \citenamefont {Erkintalo}, \citenamefont {Coen},\ and\ \citenamefont
  {Murdoch}}]{jang2015temporal}%
  \BibitemOpen
  \bibfield  {author} {\bibinfo {author} {\bibfnamefont {J.~K.}\ \bibnamefont
  {Jang}}, \bibinfo {author} {\bibfnamefont {M.}~\bibnamefont {Erkintalo}},
  \bibinfo {author} {\bibfnamefont {S.}~\bibnamefont {Coen}},\ and\ \bibinfo
  {author} {\bibfnamefont {S.~G.}\ \bibnamefont {Murdoch}},\ }\bibfield
  {title} {\bibinfo {title} {Temporal tweezing of light through the trapping
  and manipulation of temporal cavity solitons},\ }\href@noop {} {\bibfield
  {journal} {\bibinfo  {journal} {Nature communications}\ }\textbf {\bibinfo
  {volume} {6}},\ \bibinfo {pages} {7370} (\bibinfo {year} {2015})}\BibitemShut
  {NoStop}%
\bibitem [{\citenamefont {Hu}\ \emph {et~al.}(2025{\natexlab{a}})\citenamefont
  {Hu}, \citenamefont {Zhu}, \citenamefont {Yu}, \citenamefont {Wang},\ and\
  \citenamefont {Zhang}}]{hu2025dynamics}%
  \BibitemOpen
  \bibfield  {author} {\bibinfo {author} {\bibfnamefont {H.}~\bibnamefont
  {Hu}}, \bibinfo {author} {\bibfnamefont {R.}~\bibnamefont {Zhu}}, \bibinfo
  {author} {\bibfnamefont {Y.}~\bibnamefont {Yu}}, \bibinfo {author}
  {\bibfnamefont {Z.}~\bibnamefont {Wang}},\ and\ \bibinfo {author}
  {\bibfnamefont {Z.}~\bibnamefont {Zhang}},\ }\bibfield  {title} {\bibinfo
  {title} {Dynamics of kerr cavity solitons in fiber resonators driven by
  desynchronized pulses},\ }\href@noop {} {\bibfield  {journal} {\bibinfo
  {journal} {Physical Review A}\ }\textbf {\bibinfo {volume} {111}},\ \bibinfo
  {pages} {053519} (\bibinfo {year} {2025}{\natexlab{a}})}\BibitemShut
  {NoStop}%
\bibitem [{\citenamefont {Ji}\ \emph {et~al.}(2026)\citenamefont {Ji},
  \citenamefont {Ren}, \citenamefont {Yu},\ and\ \citenamefont
  {Wang}}]{4hf5-r4nc}%
  \BibitemOpen
  \bibfield  {author} {\bibinfo {author} {\bibfnamefont {J.}~\bibnamefont
  {Ji}}, \bibinfo {author} {\bibfnamefont {Z.}~\bibnamefont {Ren}}, \bibinfo
  {author} {\bibfnamefont {B.}~\bibnamefont {Yu}},\ and\ \bibinfo {author}
  {\bibfnamefont {Z.}~\bibnamefont {Wang}},\ }\bibfield  {title} {\bibinfo
  {title} {Manipulation of temporal cavity solitons in pulsed-driving kerr
  fiber resonators},\ }\href {https://doi.org/10.1103/4hf5-r4nc} {\bibfield
  {journal} {\bibinfo  {journal} {Phys. Rev. A}\ }\textbf {\bibinfo {volume}
  {114}},\ \bibinfo {pages} {023522} (\bibinfo {year} {2026})}\BibitemShut
  {NoStop}%
\bibitem [{\citenamefont {Hendry}\ \emph {et~al.}(2018)\citenamefont {Hendry},
  \citenamefont {Chen}, \citenamefont {Wang}, \citenamefont {Garbin},
  \citenamefont {Javaloyes}, \citenamefont {Oppo}, \citenamefont {Coen},
  \citenamefont {Murdoch},\ and\ \citenamefont
  {Erkintalo}}]{hendry2018spontaneous}%
  \BibitemOpen
  \bibfield  {author} {\bibinfo {author} {\bibfnamefont {I.}~\bibnamefont
  {Hendry}}, \bibinfo {author} {\bibfnamefont {W.}~\bibnamefont {Chen}},
  \bibinfo {author} {\bibfnamefont {Y.}~\bibnamefont {Wang}}, \bibinfo {author}
  {\bibfnamefont {B.}~\bibnamefont {Garbin}}, \bibinfo {author} {\bibfnamefont
  {J.}~\bibnamefont {Javaloyes}}, \bibinfo {author} {\bibfnamefont {G.-L.}\
  \bibnamefont {Oppo}}, \bibinfo {author} {\bibfnamefont {S.}~\bibnamefont
  {Coen}}, \bibinfo {author} {\bibfnamefont {S.~G.}\ \bibnamefont {Murdoch}},\
  and\ \bibinfo {author} {\bibfnamefont {M.}~\bibnamefont {Erkintalo}},\
  }\bibfield  {title} {\bibinfo {title} {Spontaneous symmetry breaking and
  trapping of temporal kerr cavity solitons by pulsed or amplitude-modulated
  driving fields},\ }\href@noop {} {\bibfield  {journal} {\bibinfo  {journal}
  {Physical Review A}\ }\textbf {\bibinfo {volume} {97}},\ \bibinfo {pages}
  {053834} (\bibinfo {year} {2018})}\BibitemShut {NoStop}%
\bibitem [{\citenamefont {Hu}\ \emph {et~al.}(2025{\natexlab{b}})\citenamefont
  {Hu}, \citenamefont {Wang}, \citenamefont {Liu}, \citenamefont {Yu},\ and\
  \citenamefont {Zhang}}]{hu2025vector}%
  \BibitemOpen
  \bibfield  {author} {\bibinfo {author} {\bibfnamefont {H.}~\bibnamefont
  {Hu}}, \bibinfo {author} {\bibfnamefont {Z.}~\bibnamefont {Wang}}, \bibinfo
  {author} {\bibfnamefont {S.}~\bibnamefont {Liu}}, \bibinfo {author}
  {\bibfnamefont {Y.}~\bibnamefont {Yu}},\ and\ \bibinfo {author}
  {\bibfnamefont {Z.}~\bibnamefont {Zhang}},\ }\bibfield  {title} {\bibinfo
  {title} {Vector kerr cavity solitons in a pulse-pumped normal-dispersion
  fiber kerr resonator},\ }\href@noop {} {\bibfield  {journal} {\bibinfo
  {journal} {Physical Review A}\ }\textbf {\bibinfo {volume} {112}},\ \bibinfo
  {pages} {023501} (\bibinfo {year} {2025}{\natexlab{b}})}\BibitemShut
  {NoStop}%
\bibitem [{\citenamefont {Garbin}\ \emph {et~al.}(2021)\citenamefont {Garbin},
  \citenamefont {Fatome}, \citenamefont {Oppo}, \citenamefont {Erkintalo},
  \citenamefont {Murdoch},\ and\ \citenamefont {Coen}}]{garbin2021dissipative}%
  \BibitemOpen
  \bibfield  {author} {\bibinfo {author} {\bibfnamefont {B.}~\bibnamefont
  {Garbin}}, \bibinfo {author} {\bibfnamefont {J.}~\bibnamefont {Fatome}},
  \bibinfo {author} {\bibfnamefont {G.-L.}\ \bibnamefont {Oppo}}, \bibinfo
  {author} {\bibfnamefont {M.}~\bibnamefont {Erkintalo}}, \bibinfo {author}
  {\bibfnamefont {S.~G.}\ \bibnamefont {Murdoch}},\ and\ \bibinfo {author}
  {\bibfnamefont {S.}~\bibnamefont {Coen}},\ }\bibfield  {title} {\bibinfo
  {title} {Dissipative polarization domain walls in a passive coherently driven
  kerr resonator},\ }\href@noop {} {\bibfield  {journal} {\bibinfo  {journal}
  {Physical Review Letters}\ }\textbf {\bibinfo {volume} {126}},\ \bibinfo
  {pages} {023904} (\bibinfo {year} {2021})}\BibitemShut {NoStop}%
\bibitem [{\citenamefont {Lucas}\ \emph {et~al.}(2025)\citenamefont {Lucas},
  \citenamefont {Xu}, \citenamefont {Wang}, \citenamefont {Oppo}, \citenamefont
  {Hill}, \citenamefont {Del’Haye}, \citenamefont {Kibler}, \citenamefont
  {Xu}, \citenamefont {Murdoch}, \citenamefont {Erkintalo} \emph
  {et~al.}}]{lucas2025polarization}%
  \BibitemOpen
  \bibfield  {author} {\bibinfo {author} {\bibfnamefont {E.}~\bibnamefont
  {Lucas}}, \bibinfo {author} {\bibfnamefont {G.}~\bibnamefont {Xu}}, \bibinfo
  {author} {\bibfnamefont {P.}~\bibnamefont {Wang}}, \bibinfo {author}
  {\bibfnamefont {G.-L.}\ \bibnamefont {Oppo}}, \bibinfo {author}
  {\bibfnamefont {L.}~\bibnamefont {Hill}}, \bibinfo {author} {\bibfnamefont
  {P.}~\bibnamefont {Del’Haye}}, \bibinfo {author} {\bibfnamefont
  {B.}~\bibnamefont {Kibler}}, \bibinfo {author} {\bibfnamefont
  {Y.}~\bibnamefont {Xu}}, \bibinfo {author} {\bibfnamefont {S.~G.}\
  \bibnamefont {Murdoch}}, \bibinfo {author} {\bibfnamefont {M.}~\bibnamefont
  {Erkintalo}}, \emph {et~al.},\ }\bibfield  {title} {\bibinfo {title}
  {Polarization faticons: Chiral localized structures in self-defocusing kerr
  resonators},\ }\href@noop {} {\bibfield  {journal} {\bibinfo  {journal}
  {Physical Review Letters}\ }\textbf {\bibinfo {volume} {135}},\ \bibinfo
  {pages} {063803} (\bibinfo {year} {2025})}\BibitemShut {NoStop}%
\bibitem [{\citenamefont {Nielsen}\ \emph {et~al.}(2019)\citenamefont
  {Nielsen}, \citenamefont {Garbin}, \citenamefont {Coen}, \citenamefont
  {Murdoch},\ and\ \citenamefont {Erkintalo}}]{nielsen2019coexistence}%
  \BibitemOpen
  \bibfield  {author} {\bibinfo {author} {\bibfnamefont {A.~U.}\ \bibnamefont
  {Nielsen}}, \bibinfo {author} {\bibfnamefont {B.}~\bibnamefont {Garbin}},
  \bibinfo {author} {\bibfnamefont {S.}~\bibnamefont {Coen}}, \bibinfo {author}
  {\bibfnamefont {S.~G.}\ \bibnamefont {Murdoch}},\ and\ \bibinfo {author}
  {\bibfnamefont {M.}~\bibnamefont {Erkintalo}},\ }\bibfield  {title} {\bibinfo
  {title} {Coexistence and interactions between nonlinear states with different
  polarizations in a monochromatically driven passive kerr resonator},\
  }\href@noop {} {\bibfield  {journal} {\bibinfo  {journal} {Physical review
  letters}\ }\textbf {\bibinfo {volume} {123}},\ \bibinfo {pages} {013902}
  (\bibinfo {year} {2019})}\BibitemShut {NoStop}%
\bibitem [{\citenamefont {Cui}\ \emph {et~al.}(2021)\citenamefont {Cui},
  \citenamefont {Zhang}, \citenamefont {Song}, \citenamefont {Huang},
  \citenamefont {Tong}, \citenamefont {Qiu},\ and\ \citenamefont
  {Liu}}]{cui2021xpm}%
  \BibitemOpen
  \bibfield  {author} {\bibinfo {author} {\bibfnamefont {Y.}~\bibnamefont
  {Cui}}, \bibinfo {author} {\bibfnamefont {Y.}~\bibnamefont {Zhang}}, \bibinfo
  {author} {\bibfnamefont {Y.}~\bibnamefont {Song}}, \bibinfo {author}
  {\bibfnamefont {L.}~\bibnamefont {Huang}}, \bibinfo {author} {\bibfnamefont
  {L.}~\bibnamefont {Tong}}, \bibinfo {author} {\bibfnamefont {J.}~\bibnamefont
  {Qiu}},\ and\ \bibinfo {author} {\bibfnamefont {X.}~\bibnamefont {Liu}},\
  }\bibfield  {title} {\bibinfo {title} {Xpm-induced vector asymmetrical
  soliton with spectral period doubling in mode-locked fiber laser},\
  }\href@noop {} {\bibfield  {journal} {\bibinfo  {journal} {Laser \& Photonics
  Reviews}\ }\textbf {\bibinfo {volume} {15}},\ \bibinfo {pages} {2000216}
  (\bibinfo {year} {2021})}\BibitemShut {NoStop}%
\bibitem [{\citenamefont {Yang}\ \emph {et~al.}(2025)\citenamefont {Yang},
  \citenamefont {Chen}, \citenamefont {Lin}, \citenamefont {Hu}, \citenamefont
  {Xu}, \citenamefont {Ma}, \citenamefont {Liang}, \citenamefont {Ling},
  \citenamefont {Xiong}, \citenamefont {Guo} \emph
  {et~al.}}]{yang2025polarization}%
  \BibitemOpen
  \bibfield  {author} {\bibinfo {author} {\bibfnamefont {Y.}~\bibnamefont
  {Yang}}, \bibinfo {author} {\bibfnamefont {X.}~\bibnamefont {Chen}}, \bibinfo
  {author} {\bibfnamefont {W.}~\bibnamefont {Lin}}, \bibinfo {author}
  {\bibfnamefont {X.}~\bibnamefont {Hu}}, \bibinfo {author} {\bibfnamefont
  {H.}~\bibnamefont {Xu}}, \bibinfo {author} {\bibfnamefont {Y.}~\bibnamefont
  {Ma}}, \bibinfo {author} {\bibfnamefont {Z.}~\bibnamefont {Liang}}, \bibinfo
  {author} {\bibfnamefont {L.}~\bibnamefont {Ling}}, \bibinfo {author}
  {\bibfnamefont {Z.}~\bibnamefont {Xiong}}, \bibinfo {author} {\bibfnamefont
  {Y.}~\bibnamefont {Guo}}, \emph {et~al.},\ }\bibfield  {title} {\bibinfo
  {title} {Polarization symmetry breaking of ghz dissipative solitons},\
  }\href@noop {} {\bibfield  {journal} {\bibinfo  {journal} {Physical Review
  Letters}\ }\textbf {\bibinfo {volume} {134}},\ \bibinfo {pages} {213803}
  (\bibinfo {year} {2025})}\BibitemShut {NoStop}%
\bibitem [{\citenamefont {Quinn}\ \emph {et~al.}(2026)\citenamefont {Quinn},
  \citenamefont {Xu}, \citenamefont {Fatome}, \citenamefont {Oppo},
  \citenamefont {Murdoch}, \citenamefont {Erkintalo},\ and\ \citenamefont
  {Coen}}]{quinn2026coherent}%
  \BibitemOpen
  \bibfield  {author} {\bibinfo {author} {\bibfnamefont {L.}~\bibnamefont
  {Quinn}}, \bibinfo {author} {\bibfnamefont {Y.}~\bibnamefont {Xu}}, \bibinfo
  {author} {\bibfnamefont {J.}~\bibnamefont {Fatome}}, \bibinfo {author}
  {\bibfnamefont {G.-L.}\ \bibnamefont {Oppo}}, \bibinfo {author}
  {\bibfnamefont {S.~G.}\ \bibnamefont {Murdoch}}, \bibinfo {author}
  {\bibfnamefont {M.}~\bibnamefont {Erkintalo}},\ and\ \bibinfo {author}
  {\bibfnamefont {S.}~\bibnamefont {Coen}},\ }\bibfield  {title} {\bibinfo
  {title} {Coherent ising machine based on polarization symmetry breaking in a
  driven kerr resonator},\ }\href@noop {} {\bibfield  {journal} {\bibinfo
  {journal} {Nature Communications}\ }\textbf {\bibinfo {volume} {17}},\
  \bibinfo {pages} {2100} (\bibinfo {year} {2026})}\BibitemShut {NoStop}%
\bibitem [{\citenamefont {Quinn}\ \emph {et~al.}(2023)\citenamefont {Quinn},
  \citenamefont {Xu}, \citenamefont {Xu}, \citenamefont {Li}, \citenamefont
  {Fatome}, \citenamefont {Murdoch}, \citenamefont {Coen},\ and\ \citenamefont
  {Erkintalo}}]{quinn2023random}%
  \BibitemOpen
  \bibfield  {author} {\bibinfo {author} {\bibfnamefont {L.}~\bibnamefont
  {Quinn}}, \bibinfo {author} {\bibfnamefont {G.}~\bibnamefont {Xu}}, \bibinfo
  {author} {\bibfnamefont {Y.}~\bibnamefont {Xu}}, \bibinfo {author}
  {\bibfnamefont {Z.}~\bibnamefont {Li}}, \bibinfo {author} {\bibfnamefont
  {J.}~\bibnamefont {Fatome}}, \bibinfo {author} {\bibfnamefont {S.~G.}\
  \bibnamefont {Murdoch}}, \bibinfo {author} {\bibfnamefont {S.}~\bibnamefont
  {Coen}},\ and\ \bibinfo {author} {\bibfnamefont {M.}~\bibnamefont
  {Erkintalo}},\ }\bibfield  {title} {\bibinfo {title} {Random number
  generation using spontaneous symmetry breaking in a kerr resonator},\
  }\href@noop {} {\bibfield  {journal} {\bibinfo  {journal} {Optics Letters}\
  }\textbf {\bibinfo {volume} {48}},\ \bibinfo {pages} {3741} (\bibinfo {year}
  {2023})}\BibitemShut {NoStop}%
\bibitem [{\citenamefont {Moroney}\ \emph {et~al.}(2022)\citenamefont
  {Moroney}, \citenamefont {Del~Bino}, \citenamefont {Zhang}, \citenamefont
  {Woodley}, \citenamefont {Hill}, \citenamefont {Wildi}, \citenamefont
  {Wittwer}, \citenamefont {S{\"u}dmeyer}, \citenamefont {Oppo}, \citenamefont
  {Vanner} \emph {et~al.}}]{moroney2022kerr}%
  \BibitemOpen
  \bibfield  {author} {\bibinfo {author} {\bibfnamefont {N.}~\bibnamefont
  {Moroney}}, \bibinfo {author} {\bibfnamefont {L.}~\bibnamefont {Del~Bino}},
  \bibinfo {author} {\bibfnamefont {S.}~\bibnamefont {Zhang}}, \bibinfo
  {author} {\bibfnamefont {M.}~\bibnamefont {Woodley}}, \bibinfo {author}
  {\bibfnamefont {L.}~\bibnamefont {Hill}}, \bibinfo {author} {\bibfnamefont
  {T.}~\bibnamefont {Wildi}}, \bibinfo {author} {\bibfnamefont {V.~J.}\
  \bibnamefont {Wittwer}}, \bibinfo {author} {\bibfnamefont {T.}~\bibnamefont
  {S{\"u}dmeyer}}, \bibinfo {author} {\bibfnamefont {G.-L.}\ \bibnamefont
  {Oppo}}, \bibinfo {author} {\bibfnamefont {M.~R.}\ \bibnamefont {Vanner}},
  \emph {et~al.},\ }\bibfield  {title} {\bibinfo {title} {A kerr polarization
  controller},\ }\href@noop {} {\bibfield  {journal} {\bibinfo  {journal}
  {Nature communications}\ }\textbf {\bibinfo {volume} {13}},\ \bibinfo {pages}
  {398} (\bibinfo {year} {2022})}\BibitemShut {NoStop}%
\bibitem [{\citenamefont {Copie}\ \emph {et~al.}(2019)\citenamefont {Copie},
  \citenamefont {Woodley}, \citenamefont {Del~Bino}, \citenamefont {Silver},
  \citenamefont {Zhang},\ and\ \citenamefont
  {Del’Haye}}]{copie2019interplay}%
  \BibitemOpen
  \bibfield  {author} {\bibinfo {author} {\bibfnamefont {F.}~\bibnamefont
  {Copie}}, \bibinfo {author} {\bibfnamefont {M.~T.}\ \bibnamefont {Woodley}},
  \bibinfo {author} {\bibfnamefont {L.}~\bibnamefont {Del~Bino}}, \bibinfo
  {author} {\bibfnamefont {J.~M.}\ \bibnamefont {Silver}}, \bibinfo {author}
  {\bibfnamefont {S.}~\bibnamefont {Zhang}},\ and\ \bibinfo {author}
  {\bibfnamefont {P.}~\bibnamefont {Del’Haye}},\ }\bibfield  {title}
  {\bibinfo {title} {Interplay of polarization and time-reversal symmetry
  breaking in synchronously pumped ring resonators},\ }\href@noop {} {\bibfield
   {journal} {\bibinfo  {journal} {Physical review letters}\ }\textbf {\bibinfo
  {volume} {122}},\ \bibinfo {pages} {013905} (\bibinfo {year}
  {2019})}\BibitemShut {NoStop}%
\bibitem [{\citenamefont {Hill}\ \emph {et~al.}(2023)\citenamefont {Hill},
  \citenamefont {Oppo},\ and\ \citenamefont {Del’Haye}}]{hill2023multi}%
  \BibitemOpen
  \bibfield  {author} {\bibinfo {author} {\bibfnamefont {L.}~\bibnamefont
  {Hill}}, \bibinfo {author} {\bibfnamefont {G.-L.}\ \bibnamefont {Oppo}},\
  and\ \bibinfo {author} {\bibfnamefont {P.}~\bibnamefont {Del’Haye}},\
  }\bibfield  {title} {\bibinfo {title} {Multi-stage spontaneous symmetry
  breaking of light in kerr ring resonators},\ }\href@noop {} {\bibfield
  {journal} {\bibinfo  {journal} {Communications Physics}\ }\textbf {\bibinfo
  {volume} {6}},\ \bibinfo {pages} {208} (\bibinfo {year} {2023})}\BibitemShut
  {NoStop}%
\end{thebibliography}%
\end{document}

\title{Supplemental Material: Competing Symmetry Breakings and Spontaneous Parity Restoration in Desynchronously Pumped Kerr Resonators}

\author{Jun Ji}
\author{Zhibin Ren}
\author{Benli Yu}
\author{Zhiqiang Wang}
\email{zhiqiangwang@ahu.edu.cn}
\affiliation{State Key Laboratory of Opto-Electronic Information Acquisition and Protection Technology \& School of Optoelectronic Science and Engineering, Anhui University, Hefei 230601, Anhui, P. R. China}
\affiliation{Information Materials and Intelligent Sensing Laboratory of Anhui Province, Anhui University, Hefei 230039, China}

\date{\today}
\maketitle

This Supplemental Material provides detailed analytical derivations and numerical specifications supporting the results presented in the main text. It is organized as follows: Appendix~A derives the effective Ginzburg-Landau potential framework from the coupled driven-dissipative Lugiato-Lefever equations (LLEs) via collective-coordinate perturbation theory. Appendix~B describes the full iterated Ikeda-map model and numerical parameters used for validation beyond the mean-field limit. Appendix~C details the mean-field reduction and dimensionless normalization mapping the physical Ikeda-map parameters onto the dimensionless coupled LLEs.

\section*{Appendix A: Collective-Coordinate Effective Potential Model}

To analytically capture the competing temporal and polarization symmetry-breaking dynamics in the driven-dissipative resonator, we reduce the infinite-dimensional coupled Lugiato-Lefever equations (LLEs),

\begin{equation}
    \frac{\partial E_{x,y}}{\partial t} = -(1+i\Delta)E_{x,y} + i\frac{d_2}{2}\frac{\partial^2 E_{x,y}}{\partial\tau^2} - v_d\frac{\partial E_{x,y}}{\partial\tau} + i\left(|E_{x,y}|^2 + \beta|E_{y,x}|^2\right)E_{x,y} + S(\tau)
    \label{eqS1}
    \tag{A1}
\end{equation}
with the localized Gaussian driving profile $S(\tau) = S_0 \exp[-\tau^2/(2\sigma^2)]$, onto a two-dimensional collective-coordinate model. Although Eq.~(\ref{eqS1}) governs a non-conservative driven-dissipative system, its quasi-steady equilibrium manifold and bifurcation topology (in the absence of oscillatory instabilities) can be faithfully captured by a joint effective Ginzburg-Landau potential $U(\tau_c,\eta)$ parameterized by the soliton temporal centroid $\tau_c(t)$ and the fractional polarization energy asymmetry $\eta(t) \in [-1,1]$.

In the quasi-steady bright cavity soliton (CS) regime, we adopt the fundamental vector soliton ansatz,

\begin{equation}
    E_{x,y}(\tau,t) = A_{x,y}(t)\,\operatorname{sech}\!\left(\frac{\tau-\tau_c(t)}{w}\right)e^{i\phi_{x,y}(t)}
    \label{eqS2}
    \tag{A2}
\end{equation}
where $\phi_{x,y}$ denote the modal phases. The modal amplitudes are parameterized as $A_x(t) = A_0\sqrt{[1+\eta(t)]/2}$ and $A_y(t) = A_0\sqrt{[1-\eta(t)]/2}$ with total amplitude $A_0$, and the characteristic soliton width $w$ and amplitude $A_0$ obey the standard anomalous-dispersion Kerr CS scaling relations ($d_2 > 0$, $\Delta > 0$),

\begin{equation}
    w \propto \sqrt{\frac{d_2}{\Delta}}, \qquad A_0 \propto \sqrt{\Delta}
    \label{eqS3}
    \tag{A3}
\end{equation}

We first evaluate the coherent driving contribution to the effective potential, which arises from the real-valued overlap integral between the localized pump field and the intracavity vector soliton:

\begin{equation}
    U_{\mathrm{pump}}(\tau_c,\eta) \propto -\mathrm{Re}\int_{-\infty}^{\infty} S(\tau)\left[E_x^*(\tau) + E_y^*(\tau)\right]d\tau
    \label{eqS4}
    \tag{A4}
\end{equation}

In the broad-pump limit ($\sigma \gg w$), $S(\tau)$ varies slowly across the soliton core and can be evaluated at $\tau_c$. Substituting Eq.~(\ref{eqS2}) into Eq.~(\ref{eqS4}) and expanding around the symmetric state ($\eta = 0$, where $\phi_x = \phi_y$) in the weak-asymmetry regime yields

\begin{equation}
    U_{\mathrm{pump}}(\tau_c,\eta) \propto -S(\tau_c)A_0 w\left(\sqrt{\frac{1+\eta}{2}} + \sqrt{\frac{1-\eta}{2}}\right) = -\sqrt{2}\,S(\tau_c)A_0 w\left(1 - \frac{1}{8}\eta^2 - \frac{5}{128}\eta^4 + \mathcal{O}(\eta^6)\right).
    \tag{A5}
\end{equation}

Because the driving envelope $S(\tau_c)$ depends explicitly on $\tau_c$, the $\eta^0$ term cannot be discarded as an irrelevant constant. Using Eq.~(\ref{eqS3}), this zeroth-order term gives the $\eta$-independent temporal tweezing potential $U_0(\tau_c) = -C_p S(\tau_c)$ (with $C_p \propto \sqrt{d_2} > 0$). Together with the linear drift tilt $-V_d\tau_c$ ($V_d \propto v_d$) induced by the asynchronous walk-off term $-v_d\partial_\tau E_{x,y}$, the translational potential $U_0(\tau_c) - V_d\tau_c$ directly recovers the established overdamped soliton drift equation $\dot{\tau}_c = v_d + \mu S'(\tau_c)$ in gradient-flow form ($\dot{\tau}_c = -\partial U/\partial\tau_c$).

\vspace{12pt}

Next, we incorporate the $\eta$-dependent polarization potential. Invariance of Eq.~(\ref{eqS1}) under mode exchange ($E_x \leftrightarrow E_y$, i.e., $\eta \leftrightarrow -\eta$) strictly forbids all odd powers of $\eta$. Retaining terms up to quartic order yields the joint effective Ginzburg-Landau potential:

\begin{equation}
    U(\tau_c,\eta) \approx U_0(\tau_c) - V_d\tau_c + A(\tau_c)\eta^2 + B(\tau_c)\eta^4
    \label{eqS6}
    \tag{A6}
\end{equation}
where $B(\tau_c) = \kappa_B S(\tau_c) + C_0 > 0$ ($\kappa_B, C_0 > 0$) ensures global boundedness. The quadratic bifurcation coefficient $A(\tau_c)$ encapsulates the competition between linear cavity stabilization and pump-sustained inter-modal cross-phase modulation (XPM) repulsion. Specifically, the linear stabilizing contribution from cavity damping and detuning scales with the total soliton energy, $\int_{-\infty}^{\infty}(|E_x|^2 + |E_y|^2)\,d\tau \propto A_0^2 w \propto \sqrt{d_2\Delta}$. Conversely, the destabilizing inter-modal XPM repulsion (active for $\beta > 1$) is sustained by the local pump-soliton overlap and scales as $-(\beta-1)S(\tau_c)A_0 w \propto -(\beta-1)\sqrt{d_2}\,S(\tau_c)$, where the $\sqrt{\Delta}$ and $1/\sqrt{\Delta}$ factors cancel identically. Combining both contributions yields the explicit scaling of $A(\tau_c)$:

\begin{equation}
    A(\tau_c) = \kappa_{\mathrm{lin}}\sqrt{d_2\Delta} - \kappa_{\mathrm{NL}}(\beta-1)\sqrt{d_2}\,S_0\exp\!\left(-\frac{\tau_c^2}{2\sigma^2}\right)
    \label{eqS7}
    \tag{A7}
\end{equation}
where $\kappa_{\mathrm{lin}}, \kappa_{\mathrm{NL}} > 0$ are dimensionless effective constants determined by the $\operatorname{sech}$-profile overlap integrals. Setting $A(\tau_c) = 0$ defines the critical local pump threshold for polarization symmetry breaking (PSB):

\begin{equation}
    S_{\mathrm{th}} \equiv S_0\exp\!\left(-\frac{\tau_c^2}{2\sigma^2}\right) = \frac{\kappa_{\mathrm{lin}}}{\kappa_{\mathrm{NL}}(\beta-1)}\sqrt{\Delta}
    \label{eqS8}
    \tag{A8}
\end{equation}
which exhibits the canonical $\sqrt{\Delta}$ scaling, diverges as $\beta \to 1^+$ (confirming the absence of PSB for $\beta \le 1$), and determines the critical temporal offset $\tau_{\mathrm{th}} = \sigma\sqrt{2\ln(S_0/S_{\mathrm{th}})}$.

\vspace{12pt}

Finally, minimizing Eq.~(\ref{eqS6}) with respect to $\eta$ ($\partial U/\partial\eta = 0$) yields $\eta_{\mathrm{eq}}^2 = -A(\tau_c)/[2B(\tau_c)]$ for $|\tau_c| < \tau_{\mathrm{th}}$ ($A < 0$) and $\eta_{\mathrm{eq}} = 0$ for $|\tau_c| \ge \tau_{\mathrm{th}}$ ($A \ge 0$). Substituting $\eta_{\mathrm{eq}}$ back into Eq.~(\ref{eqS6}) gives the reduced one-dimensional effective potential governing the soliton centroid:

\begin{equation}
    U_{\mathrm{eff}}(\tau_c) =
    \begin{cases}
    -C_p S(\tau_c) - V_d\tau_c - \dfrac{A(\tau_c)^2}{4B(\tau_c)}, & |\tau_c| < \tau_{\mathrm{th}}\ \text{(PSB/HSB)}, \\[10pt]
    -C_p S(\tau_c) - V_d\tau_c, & |\tau_c| \ge \tau_{\mathrm{th}}\ \text{(pure TSB)}
    \end{cases}
    \tag{A9}
\end{equation}

The steady-state temporal equilibrium $\tau_c^{\mathrm{eq}}$ is determined by the mechanical force-balance condition $\partial U_{\mathrm{eff}}/\partial\tau_c = 0$. In the pure TSB regime ($|\tau_c| \ge \tau_{\mathrm{th}}$, $\eta = 0$), this condition reduces to

\begin{equation}
    V_d = -C_p\left.\frac{dS(\tau_c)}{d\tau_c}\right|_{\tau_c^{\mathrm{eq}}} = C_p S_0\frac{\tau_c^{\mathrm{eq}}}{\sigma^2}\exp\!\left[-\frac{(\tau_c^{\mathrm{eq}})^2}{2\sigma^2}\right]
    \label{eqS10}
    \tag{A10}
\end{equation}
which admits a stably locked soliton position on the driving pulse flank up to the maximum trapping drift $V_{d,\max} = C_p S_0/(\sigma\sqrt{e})$ at $\tau_c^{\mathrm{eq}} = \sigma$. In contrast, within the PSB/HSB regime ($|\tau_c| < \tau_{\mathrm{th}}$, $\eta \neq 0$), spontaneous polarization breaking lowers the potential by the condensation energy $-A(\tau_c)^2/[4B(\tau_c)]$, carving a deep central trapping well on top of $U_0(\tau_c)$. As $V_d$ increases, the inner HSB minimum shifts along the pulse flank until its restoring barrier vanishes via a saddle-node bifurcation, triggering an abrupt transition into the outer pure TSB potential well governed by Eq.~(\ref{eqS10}) with restored polarization parity.

Since the inner HSB well terminates at $|\tau_c| = \tau_{\mathrm{th}}$ where the polarization condensation energy vanishes, evaluating the restoring force-balance condition Eq.~(\ref{eqS10}) at the threshold boundary $\tau_c^{\mathrm{eq}} = \tau_{\mathrm{th}} = \sigma\sqrt{2\ln(S_0/S_{\mathrm{th}})}$ [where $S(\tau_{\mathrm{th}}) = S_{\mathrm{th}}$] yields the analytical critical drift for the HSB-to-TSB transition:

\begin{equation}
    V_{d,c}(S_0) = \frac{C_p S_{\mathrm{th}}}{\sigma}\sqrt{2\ln\!\left(\frac{S_0}{S_{\mathrm{th}}}\right)}.
    \tag{A11}
\end{equation}

Mapping back to the physical Ikeda-map parameters via $S_0 \propto \sqrt{P_0}$ and $V_d = c_V v_d = c_V(2\kappa L/\alpha T_0)$ (Appendix~C), the critical desynchronization threshold scales with pump peak power $P_0$ as

\begin{equation}
    \kappa_c(P_0) = \frac{\alpha T_0 C_p S_{\mathrm{th}}}{2 c_V L\sigma}\sqrt{\ln\!\left(\frac{P_0}{P_{0,\mathrm{th}}}\right)}
    \label{eqS12}
    \tag{A12}
\end{equation}
where $P_{0,\mathrm{th}} = \alpha^3 S_{\mathrm{th}}^2/(8\gamma L\theta)$ is the zero-desynchronization PSB threshold power. Equation~(\ref{eqS12}) defines the analytical phase boundary plotted in Fig.~3(b) of the main text.

\section*{Appendix B: Numerical Model of the Iterated Ikeda Map}

To validate the analytical predictions beyond the mean-field approximation of the LLE, we perform full iterative roundtrip simulations based on the Ikeda map, which resolves the complete spatiotemporal evolution of the intracavity fields without spatial averaging. During each roundtrip, the propagation of the two orthogonally polarized field components $E_{x,y}(Z,T)$ through the passive nonlinear fiber is governed by the coupled nonlinear Schr\"odinger equations (NLSEs):

\begin{equation}
    \frac{\partial E_{x,y}}{\partial Z} = -\kappa\frac{\partial E_{x,y}}{\partial T} - i\frac{\beta_2}{2}\frac{\partial^2 E_{x,y}}{\partial T^2} + i\gamma\left(|E_{x,y}|^2 + B|E_{y,x}|^2\right)E_{x,y}
    \label{eqB1}
    \tag{B1}
\end{equation}
where $Z$ is the longitudinal spatial coordinate along the fiber and $T$ is the fast-time coordinate. The walk-off parameter $\kappa$ quantifies the temporal shift per unit propagation length arising from the desynchronization between the pump repetition rate and the cavity free spectral range (FSR), serving as the physical counterpart to $v_d$. Here, $\beta_2$ and $\gamma$ denote the group-velocity dispersion (GVD) and Kerr nonlinearity coefficients, respectively, and $B$ is the XPM-to-SPM coupling ratio.

After propagating through the fiber length $L$, the intracavity fields for the subsequent $(m+1)$-th roundtrip are updated by the discrete boundary condition at the input coupler:
\begin{equation}
    E_{x,y}^{m+1}(0,T) = \sqrt{\theta}\,E_{\mathrm{in},x,y}(T) + \sqrt{1-\alpha}\,E_{x,y}^{m}(L,T)e^{-i\delta_{x,y}}
    \label{eqB2}
    \tag{B2}
\end{equation}
where $m$ is the roundtrip index, $\theta$ is the power coupling ratio, $\alpha$ is the total roundtrip power loss (including both intrinsic fiber loss and coupler out-coupling loss), and $\delta_{x,y}$ are the roundtrip linear phase detunings. The resonator is symmetrically driven by equal-amplitude localized Gaussian pump pulses:

\begin{equation}
    E_{\mathrm{in},x,y}(T) = \sqrt{P_0}\exp\!\left(-\frac{T^2}{2\tau_0^2}\right)
    \label{eqB3}
    \tag{B3}
\end{equation}
with single-polarization peak pump power $P_0$ and pulse width $\tau_0$.

Equations~(\ref{eqB1}) and (\ref{eqB2}) are integrated numerically using the standard split-step Fourier method. In our simulations, the cavity consists of a $L = 100~\mathrm{m}$ standard single-mode fiber (SMF) with $\beta_2 = -22~\mathrm{ps^2/km}$, $\gamma = 1.2~\mathrm{W^{-1}km^{-1}}$, $B = 2$, $\theta = 0.05$, $\alpha = 0.28$, and $\tau_0 = 15~\mathrm{ps}$. To ensure symmetric driving of both polarization modes, we set identical detunings $\delta_x = \delta_y = 0.2$ and scan the pump peak power within $P_0 \in [0.082, 0.094]~\mathrm{W}$. Simulations are initialized from weak random noise with a maximum peak power of $0.01~\mathrm{W}$ superimposed on both polarization components.

\section*{Appendix C: Mapping from the Ikeda Map to the Dimensionless Coupled LLEs}

To bridge the numerical Ikeda-map simulations [Eqs.~(\ref{eqB1}) and (\ref{eqB2})] with the analytical model in Appendix~A and the main text, we perform the standard mean-field reduction. 

In the good-cavity limit where $\alpha \ll 1$, $|\delta_{x,y}| \ll 1$, and intracavity field variations per roundtrip are small, we expand the boundary condition Eq.~(B2) to first order:

\begin{equation}
    \sqrt{1-\alpha}\,e^{-i\delta_{x,y}} \approx 1 - \frac{\alpha}{2} - i\delta_{x,y}
    \label{eqC1}
    \tag{C1}
\end{equation}
Integrating Eq.~(\ref{eqB1}) over one cavity roundtrip ($Z \in [0,L]$) to first order yields $E_{x,y}^{m}(L,T) \approx E_{x,y}^{m}(0,T) + L\partial_Z E_{x,y}^{m}(0,T)$. Combining this with Eq.~(\ref{eqB2}) and replacing the roundtrip difference $E_{x,y}^{m+1}(0,T) - E_{x,y}^{m}(0,T)$ with the continuous slow-time derivative $t_R\partial E_{x,y}/\partial t_s$ (assuming symmetric detuning $\delta_x = \delta_y = \delta_0$ and symmetric pumping $E_{\mathrm{in},x} = E_{\mathrm{in},y} = E_{\mathrm{in}}$) gives the dimensional coupled LLEs:

\begin{equation}
    t_R\frac{\partial E_{x,y}}{\partial t_s} = \left[-\frac{\alpha}{2} - i\delta_0 - \kappa L\frac{\partial}{\partial T} - i\frac{\beta_2 L}{2}\frac{\partial^2}{\partial T^2} + i\gamma L\left(|E_{x,y}|^2 + B|E_{y,x}|^2\right)\right]E_{x,y} + \sqrt{\theta}\,E_{\mathrm{in}}(T)
    \label{eqC2}
    \tag{C2}
\end{equation}
where $t_s = m t_R$ is the continuous slow evolution time ($t_R$ is the cavity roundtrip time) and $\kappa L$ is the pump-cavity temporal walk-off per roundtrip.

Introducing a reference fast-time scale $T_0$, identifying the XPM coupling ratio $\beta \equiv B$, and applying the rescaling transformations
\begin{align}
    t &= \frac{\alpha}{2t_R}t_s = \frac{\alpha}{2}m, & \tau &= \frac{T}{T_0}, & \tilde{E}_{x,y} &= E_{x,y}\sqrt{\frac{2\gamma L}{\alpha}}, & \Delta &= \frac{2\delta_0}{\alpha}, \notag\\
    v_d &= \frac{2\kappa L}{\alpha T_0}, & d_2 &= -\frac{2\beta_2 L}{\alpha T_0^2}, & S(\tau) &= E_{\mathrm{in}}(T)\sqrt{\frac{8\gamma L\theta}{\alpha^3}}, & &
    \label{eqC3}
    \tag{C3}
\end{align}
directly casts Eq.~(\ref{eqC2}) into the dimensionless coupled LLEs Eq.~(\ref{eqS1}) (after dropping the tildes). Choosing the characteristic dispersion time scale $T_0 = \sqrt{2|\beta_2|L/\alpha} \approx 3.9641~\mathrm{ps}$ simplifies the dimensionless dispersion to $d_2 = -\operatorname{sgn}(\beta_2) = +1$ in the anomalous dispersion regime ($\beta_2 < 0$).

Under this scaling, the Gaussian pump pulse Eq.~(\ref{eqB3}) with physical width $\tau_0$ and peak power $P_0$ maps to the normalized pulse width $\sigma$ and dimensionless driving amplitude $S_0$ as

\begin{equation}
    \sigma = \frac{\tau_0}{T_0} = \tau_0\sqrt{\frac{\alpha}{2|\beta_2|L}}, \qquad S_0 = \sqrt{\frac{8\gamma L\theta P_0}{\alpha^3}}
    \label{eqC4}
    \tag{C4}
\end{equation}
which fully specify $S(\tau) = S_0\exp[-\tau^2/(2\sigma^2)]$ in Eq.~(\ref{eqS1}). For the effective potential landscape plotted in Fig.~1(b)--(d) of the main text ($P_0 = 0.09~\mathrm{W}$, $\delta_0 = 0.2$), Eqs.~(\ref{eqC3}) and \ref{eqC4}) yield $\sigma = 3.7839$, $S_0 = 0.4436$, $\Delta = 1.4286$, $d_2 = 1$, and $\beta = 2$. Because the full Ikeda-map simulations operate with finite roundtrip loss ($\alpha = 0.28$) and a non-zero pulsed background beyond the strict high-finesse limit, the dimensionless overlap constants are calibrated against the numerical benchmark as $V_d = c_V v_d = 0.225 v_d$, $\kappa_{\mathrm{lin}} = 0.3275$, $\kappa_{\mathrm{NL}} = 1.0$, $\kappa_B = 0.06$, $C_p = 0.088$, and $C_0 = 0.026$, preserving the exact analytical scaling laws across the entire parameter space.